\PassOptionsToPackage{table}{xcolor}
\documentclass[10pt,journal,compsoc]{IEEEtran}

\newcommand{\gemma}{\textsc{Gemma-3-4B}}
\newcommand{\llama}{\textsc{Llama-3.2-3B}}
\newcommand{\phimodel}{\textsc{Phi-4-14B}}

\newcommand{\gemini}{\textsc{Gemini-3-Pro-Preview}}
\newcommand{\gpt}{\textsc{ChatGPT 5.2 Thinking}}
\newcommand{\sensei}{\textsc{Manual Test Sensei}}

\usepackage{xcolor}
\definecolor{responseblue}{RGB}{0,0,0}

\newcommand{\numTestCases}{100}
\newcommand{\numStepsTotal}{2,618}
\newcommand{\numStepsManual}{103}
\newcommand{\numTestsManual}{15}
\newcommand{\amostraManual}{103}

\newcommand{\precision}{0.92}
\newcommand{\recall}{0.91}
\newcommand{\fscore}{0.91}

\newcommand{\precisionGPT}{0.84}
\newcommand{\recallGPT}{0.80}
\newcommand{\fscoreGPT}{0.82}

\newcommand{\precisionSensei}{0.70}
\newcommand{\recallSensei}{0.62}
\newcommand{\fscoreSensei}{0.66}

\usepackage{tikz}
\usetikzlibrary{calc,positioning}
\definecolor{overAB}{RGB}{232,218,170}  
\definecolor{overAC}{RGB}{140,185,175}  
\definecolor{overBC}{RGB}{205,215,155}  

\definecolor{overABC}{RGB}{165,155,105} 
\usepackage{pgfplots}
\pgfplotsset{compat=1.18}
\usepgfplotslibrary{statistics}
\usepackage[most]{tcolorbox}

\newtcolorbox{geminioutput}{
  enhanced,
  boxrule=0.4pt,
  arc=2pt,
  left=6pt,
  right=6pt,
  top=4pt,
  bottom=4pt,
  colback=cyan!4,        
  colframe=cyan!35,      
  fontupper=\itshape
}
\usepackage{xcolor}
\usepackage{listings}
\usepackage[most]{tcolorbox}

\lstdefinestyle{xmlsnippet}{
  language={},
  basicstyle=\small\ttfamily\color{black},
  numbers=left,
  numberstyle=\tiny\color{black},
  frame=single,
  breaklines=true,
  breakindent=0pt,
  prebreak={},
  postbreak={},
  columns=fullflexible,
  keepspaces=true,
  showspaces=false,
  showtabs=false,
  showstringspaces=false,
  keywordstyle=\color{black},
  commentstyle=\color{black},
  stringstyle=\color{black},
  identifierstyle=\color{black},
  literate=
    {<dl>}{{\textcolor{blue}{<dl>}}}4
    {</dl>}{{\textcolor{blue}{</dl>}}}5
    {<dt>}{{\textcolor{blue}{<dt>}}}4
    {</dt>}{{\textcolor{blue}{</dt>}}}5
    {<dd>}{{\textcolor{blue}{<dd>}}}4
    {</dd>}{{\textcolor{blue}{</dd>}}}5
}

\newtcblisting{stepexample}{
  enhanced,
  listing only,
  listing engine=listings,
  listing options={
    style=xmlsnippet,
    numbers=none,
    frame=none,
    xleftmargin=0pt,
    xrightmargin=0pt,
    numbersep=0pt,
    aboveskip=0pt,
    belowskip=0pt,
    breaklines=true,
    breakindent=0pt,
    prebreak={},
    postbreak={},
    showspaces=false,
    showtabs=false,
    showstringspaces=false
  },
  boxrule=0.4pt,
  arc=0pt,
  boxsep=0pt,
  left=3pt,
  right=3pt,
  top=4pt,
  bottom=4pt,
  colback=white,
  colframe=black!65
}

\usepackage{hologo}
\usepackage{longtable,booktabs}
\usepackage{moreverb}
\usepackage{subcaption}
\usepackage{textcomp}
\usepackage{multirow}
\usepackage{multicol}
\usepackage{url}
\usepackage{booktabs}
\usepackage{framed}
\usepackage{xspace}
\usepackage{pdflscape}
\usepackage{array}

\usepackage{listingcustom}
\usepackage{float}

\usepackage{graphicx}
\usepackage[misc,geometry]{ifsym} 
\usepackage{fontawesome}
\usepackage{academicons}
\usepackage{color}
\usepackage{amsmath}
\usepackage{hyperref} 
\usepackage{aas_macros}
\usepackage[bottom]{footmisc}
\usepackage{soul}
\usepackage{tcolorbox}
\tcbuselibrary{breakable}
\tcbset{
  width=0.45\textwidth,
  halign=justify,
  center,
  breakable,
  colback=white    
}
\usepackage{pgfplots}
\pgfplotsset{compat=1.18}
\definecolor{light-gray}{gray}{0.95}

\definecolor{darkgreen}{rgb}{0.17, 0.48, 0.10}

\definecolor{darkblue}{rgb}{0.37,0.48,0.13}
\definecolor{orcidlogo}{rgb}{0.37,0.48,0.13}
\definecolor{unilogo}{rgb}{0.16, 0.26, 0.58}
\definecolor{maillogo}{rgb}{0.58, 0.16, 0.26}

\hypersetup{colorlinks,breaklinks,
            linkcolor=darkblue,urlcolor=darkblue,
            anchorcolor=darkblue,citecolor=darkblue}
\hypersetup{colorlinks,citecolor=blue,linkcolor=blue,urlcolor=blue}

\definecolor{shadecolor}{rgb}{0.75, 0.75, 0.75}
\hypersetup{colorlinks=true, linkcolor=blue, urlcolor=blue}

\usepackage{mdframed}
\usepackage{totcount}
\newtotcounter{IconsCounter}
\usepackage{balance}

\begin{document}
\title{An Empirical Study of Gemini 3 for Detecting Natural Language Test Smells in Manual Test Cases}

\author{
    \IEEEauthorblockN{
        Keila Lucas\IEEEauthorrefmark{3}, Rohit Gheyi\IEEEauthorrefmark{3}, Márcio Ribeiro\IEEEauthorrefmark{2}, Fabio Palomba\IEEEauthorrefmark{1}, Luana Martins\IEEEauthorrefmark{1}, Elvys Soares\IEEEauthorrefmark{4}\\
    }
    \IEEEauthorblockA{
        \IEEEauthorrefmark{3}UFCG, Brazil\\ keila.santos@copin.ufcg.edu.br and rohit@dsc.ufcg.edu.br\\
    }
    \IEEEauthorblockA{
        \IEEEauthorrefmark{2}UFAL, Brazil\\ marcio@ic.ufal.br\\
    }
    \IEEEauthorblockA{
        \IEEEauthorrefmark{1}University of Salerno, Italy\\ fpalomba@unisa.it and lalmeidamartins@unisa.it \\
    }
    \IEEEauthorblockA{
        \IEEEauthorrefmark{4}IFAL, Brazil\\ elvys.soares@ifal.edu.br\\
    }
}

\maketitle

\begin{abstract}
Manual testing, in which testers follow natural language instructions to validate system behavior, remains essential for uncovering issues that are difficult to capture with automation. However, manual test cases often contain \emph{test smells}, quality issues such as ambiguity, redundancy, or missing checks that reduce reliability, maintainability, and reproducibility. Existing detection approaches largely depend on manually engineered rules and thus struggle to generalize and scale across heterogeneous test suites.
In our previous work, we assessed the feasibility of using Small Language Models (SLMs) for test smell detection by evaluating \gemma{}, \llama{}, and \phimodel{} on test steps from 143 real-world Ubuntu test cases, covering seven smell types. \phimodel{} achieved the best performance. In this article, we investigate whether a contemporary Large Language Model (\gemini{}) available at the time of the study can identify test smells in natural language manual test cases using a prompt-based, whole-test-case analysis strategy. Unlike approaches that analyze individual test steps in isolation, our approach evaluates complete test cases, enabling the model to consider relationships and dependencies among test steps. We evaluate the approach on \numTestCases{} Ubuntu test cases covering seven test smell types and compare its performance against previously evaluated SLMs, including \gemma{}, \llama{}, and \phimodel{}. Our results show that \gemini{} outperforms the SLMs, while producing actionable explanations that can help practitioners revise manual test cases for greater clarity and consistency. We also find that test smells are pervasive in practice, with nearly one detected test smell per step on average, highlighting the need for scalable and automated quality support for manual testing artifacts.
\end{abstract}

\begin{IEEEkeywords}
Test Smell, LLM, Manual Test Case.
\end{IEEEkeywords}

\section{Introduction}
\label{sec:introduction}

Manual testing is a widely used software verification technique in which testers execute predefined steps to uncover defects in the system~\cite{hauptmann2016,SoaresAORGSMSFB23}. In practice, these steps are typically documented in natural language and describe how to interact with the system and what to observe as evidence of correct behavior. Even in organizations with mature automated testing pipelines, manual testing remains essential for scenarios that are difficult to automate or validate reliably with scripts, such as exploratory workflows, complex user interactions, and environment-dependent behaviors~\cite{Hauptmann2013}.

Despite its importance, this testing approach often suffers from quality and reliability issues. Manual test descriptions are frequently authored as operational documentation rather than as carefully engineered artifacts, and they may be created and updated outside formal software engineering practices~\cite{Hauptmann2013}. As a consequence, manual tests often contain \emph{test smells}; that is, quality problems that make tests harder to understand, maintain, and execute consistently. Prior work reports a broad range of such issues, including ambiguity, redundancy, unnecessary complexity, missing checks, and checks that are too weak or underspecified~\cite{Hauptmann2013,juhnke2021clustering,junior2021,junior2020survey,SoaresAORGSMSFB23}. For instance, the \textit{Ambiguous Test} smell occurs when a step provides vague instructions, such as \textit{``Select a keyboard layout and click Continue''}, without specifying which layout should be chosen or what constitutes the correct selection. In this case, different testers may legitimately interpret the same step in different ways, leading to inconsistent test execution and reduced confidence in the test results~\cite{emse-2025}.

Improving the clarity and specificity of manual tests is challenging in practice. Writing high-quality natural language instructions requires domain knowledge, testing expertise, and careful attention to phrasing, yet test suites evolve continuously with the system. As a result, manual test repositories can accumulate quality issues over time, and systematically identifying them becomes a recurring maintenance task. Recent empirical studies have shown that smells in manual tests are not merely cosmetic, but can hinder execution and reduce efficiency. For example, Soares~\emph{et al.}~report evidence that test smells can negatively affect the efficiency of manual test execution~\cite{emse-2025}. Other studies highlight recurring problems such as ambiguous descriptions, excessive complexity, actions without checks, and inadequate checks, which can undermine the effectiveness of manual testing efforts~\cite{SoaresAORGSMSFB23,peixoto2025effectiveness}. These findings reinforce the relevance of practical techniques that help teams continuously assess and improve the quality of manual tests at scale.

However, existing automated support remains limited. Many approaches for detecting smells in natural language rely on manually engineered rules, keyword lists, or heuristics tailored to specific writing conventions. While effective in narrow settings, such approaches typically require substantial implementation effort, may be brittle under vocabulary or style variation, and do not scale well across heterogeneous test suites or evolving projects. This limitation is particularly problematic because manual tests are often maintained by different contributors over long periods, which naturally leads to diverse writing styles and inconsistent levels of detail. A previous approach proposed a tool-based method using natural language processing to detect manual test smells~\cite{SoaresAORGSMSFB23}. However, it does not perform richer semantic analysis.

In parallel, the rapid evolution of foundation models has opened new opportunities for analyzing and improving natural language artifacts in software engineering, providing richer semantic analysis. In our previous study~\cite{keila-sbes-2025}, we investigated the feasibility of using Small Language Models (SLMs) to detect seven types of test smells in manual test descriptions~\cite{Aranda-ease2024}. We evaluated \gemma{}, \llama{}, and \phimodel{} on a dataset of 261 test steps extracted from 143 manual test cases of the Ubuntu operating system, focusing on smells related to ambiguity, incomplete checks, unnecessary complexity, and redundancy. The results show that SLMs can detect test smells in practice, with \phimodel{} achieving the best overall performance and the most stable behavior across temperature settings.

In this article, we investigate whether contemporary Large Language Models (LLMs), available at the time of the study, can provide scalable support for identifying test smells in natural language manual tests. Our key idea is to analyze complete test cases rather than isolated test-case steps. Whole-test-case analysis allows the model to leverage step-level context, capture dependencies between actions and verifications, and detect smells that only become evident when considering the surrounding steps. This setting is closer to how practitioners read and review manual tests, and it enables explanations that can be grounded in the broader flow of the procedure. We instantiate this idea using \gemini{} with an improved prompting strategy and evaluate it on \numTestCases{} Ubuntu test cases, covering a substantially larger set of test-case steps and seven test smell types.

We evaluate \gemini{} under this whole-test-case setting. Our results show that \gemini{} consistently outperforms the SLM baselines and generates explanations that are actionable for revising manual tests. Beyond predictive performance, we provide evidence that test smells are pervasive in practice, with nearly one test smell per test-case step on average. This highlights that test smell detection is not a corner case, but rather a recurring maintenance concern in real test repositories.

These findings have practical implications for both researchers and practitioners. For practitioners, they suggest that LLM-based assistants can support systematic review of manual tests at scale, improving clarity and consistency while reducing manual inspection effort. For researchers, our results motivate further work on context-aware test quality analysis, robust prompting strategies, and hybrid approaches that combine model-based reasoning with lightweight checks and tooling integration. All data and artifacts are publicly available online~\cite{artefatos}.

This article is organized as follows. Section~\ref{sec:test-smells} describes the natural language test smells considered in this study. Section~\ref{sec:methodology} details our methodology. Section~\ref{sec:results} presents and discusses the results. Section~\ref{sec:related} situates our work with respect to related studies. Finally, Section~\ref{sec:conclusions} concludes the article and outlines implications and future work.

\section{Natural Language Test Smells}
\label{sec:test-smells}

Manual test cases are often written as natural language procedures that guide testers through actions and expected outcomes. Prior work shows that these artifacts frequently suffer from \emph{natural language test smells}, that is, recurring quality issues that reduce clarity, consistency, and maintainability, and can ultimately compromise the reliability of manual testing outcomes~\cite{Hauptmann2013,SoaresAORGSMSFB23,emse-2025,peixoto2025effectiveness,Aranda-ease2024}. In this article, we focus on seven smells that are both widely discussed in the literature and directly operationalized in our prompting strategy: Ambiguous Test, Conditional Test, Eager Action, Misplaced Action, Misplaced Precondition, Misplaced Verification, and Unverified Action~\cite{Hauptmann2013,SoaresAORGSMSFB23,Aranda-ease2024}. These smells are particularly relevant because they can be identified from the text of test steps and their role in the test structure, and they directly impact test determinism and the ability of practitioners to follow and interpret tests consistently~\cite{emse-2025}.

This section provides the background necessary to understand our study. We first introduce the structure of Ubuntu-style manual test cases in Section~\ref{s:ubuntu}. We then present the seven test smells analyzed in this work: Ambiguous Test (Section~\ref{s:ambiguous-test}), Conditional Test (Section~\ref{s:conditional-test}), Eager Action (Section~\ref{s:eager-action}), Misplaced Action (Section~\ref{s:misplaced-action}), Misplaced Precondition (Section~\ref{s:misplaced-precondition}), Misplaced Verification (Section~\ref{s:misplaced-verification}), and Unverified Action (Section~\ref{s:unverified-action}). Finally, Section~\ref{practical-relevance} discusses the practical relevance of these smells using evidence from Ubuntu manual-test maintenance. These sections provide the basis for both the manual annotation process and the prompting strategy evaluated in this work.

\subsection{Ubuntu-Style Manual Tests}
\label{s:ubuntu}

In this article, we focus on manual test cases from the Ubuntu ecosystem~\cite{ubuntu_manual_tests}.
Ubuntu is one of the most widely used Linux distributions across desktop, cloud, and IoT settings.\footnote{\url{https://canonical.com/blog/ubuntu-20-04-survey-results}}
We chose Ubuntu for three complementary reasons.
First, Ubuntu provides a large, publicly accessible corpus of human-authored manual test cases, curated in a consistent, semi-structured format, which supports transparent and reproducible experimentation (e.g., the public \texttt{ubuntu-manual-tests} repository).
Second, restricting the study to a single ecosystem reduces confounding variation (e.g., domain-specific terminology, documentation conventions, and test templates), helping us isolate language-level test smells rather than differences in writing styles across projects.
Third, Ubuntu test cases are authored and maintained by a broad community and cover diverse end-user workflows, which mitigates single-author bias and strengthens ecological validity.
We view this scope as a conservative first step; evaluating additional domains (e.g., mobile apps, web systems, or API-oriented documentation) is left as future work.

Ubuntu-style manual tests commonly separate what the tester \emph{does} from what the tester \emph{checks}. In Ubuntu’s manual test specifications, the procedure is often encoded using an HTML-like list structure. A test case is represented as a \texttt{<dl>} block, where each \texttt{<dt>} element denotes an action step and each \texttt{<dd>} element denotes a verification step associated with the most recent action sequence. 
When a test case requires a prerequisite state or setup condition before execution, this information should be specified in the precondition section of the manual test case, before the executable procedural steps encoded with \texttt{<dt>} and \texttt{<dd>}. In our dataset of manual tests, however, preconditions are not explicitly marked with a dedicated HTML tag, unlike the \texttt{<dt>} and \texttt{<dd>} elements. Instead, the setup description or prerequisite conditions usually appear separately before the executable steps, either as free text or as an introductory paragraph preceding the first test action. Therefore, preconditions describe the state that must already hold before the tester begins execution, while the \texttt{<dt>} and \texttt{<dd>} entries represent, respectively, the tester’s actions and the expected observations during execution.

This representation makes the intended role of each step explicit and supports a structured reading of the test: actions describe interactions with the system under test, while verifications describe the expected observable outcomes.

This separation is crucial for two reasons. First, several smells are defined by \emph{role mismatches}, such as an action written as a verification or a verification written as an action. Second, the absence of an expected outcome after an action is itself a smell, because it leaves the tester without guidance about what constitutes correct behavior. In our study, we therefore interpret \texttt{<dt>} entries as \emph{action} steps and \texttt{<dd>} entries as \emph{verification} steps, using this structural information both to contextualize smell definitions and to enable the detection of smells that depend on the presence or absence of verifications.

\noindent\textbf{Example.} Listing~\ref{lst:ubuntu-dl}\footnote{\href{https://github.com/easy-software-ufal/manual-test-sensei/blob/main/testcases/image/1442_Mythbuntu\%20Frontend}{Manual test case \#1442\_Mythbuntu Frontend (image)}} illustrates the Ubuntu-style encoding, in which preconditions are listed first, followed by action and verification elements encoded as \texttt{<dt>} and \texttt{<dd>}, respectively. A verification may state the expected outcome of multiple preceding actions.

\begin{figure*}[t]
\centering
\begin{lstlisting}[style=xmlsnippet,caption={Ubuntu manual test case.},label={lst:ubuntu-dl}]
The testcase tests the installation and basic functionality of a Frontend. 
You will need a MythTV backend in the network to do this testcase completely.

<dl>
    <dt>Go through the Ubiquity install as you would any other install. [...]</dt>
    [...]
    <dt>Once the machine boots up, it should boot into the frontend</dt>
    <dt>Select "Watch TV". This will only work if you have a backend already in the network</dt>
        <dd>The demo video should start playing</dd>
</dl>
\end{lstlisting}
\end{figure*}

\subsection{Ambiguous Test}
\label{s:ambiguous-test}

\noindent\textbf{Definition.} Ambiguous Test occurs when a step is vague or underspecified, leaving room for multiple interpretations and preventing an objective pass or fail decision~\cite{Hauptmann2013,SoaresAORGSMSFB23,Aranda-ease2024}. Ambiguity often stems from indefinite determiners, unclear references, subjective qualifiers, or comparative claims without measurable criteria.

\noindent\textbf{Why it is harmful.} Ambiguity makes manual testing non-deterministic. Two testers can follow the same step and legitimately choose different inputs, paths, or interpretations of success, producing inconsistent outcomes and lowering confidence in the test suite~\cite{Hauptmann2013,emse-2025}.

\noindent\textbf{Examples.} The following step is ambiguous because it does not specify which layout should be selected, which can lead testers to make different choices.

\begin{stepexample}
<dt>Select a keyboard layout and click Continue</dt>
\end{stepexample}

\noindent In the next example, the expected outcome is described using subjective language, and the test does not define concrete observable criteria for correctness.
\begin{stepexample}
<dt>Press 'Apply'. Verify that the new resolution is set correctly and that the screen is still visible.</dt>
\end{stepexample}

\subsection{Conditional Test}
\label{s:conditional-test}
\noindent\textbf{Definition.} Conditional Test appears when a step embeds branching logic, such as conditions expressed with markers like if, when, unless, or otherwise~\cite{Hauptmann2013,Aranda-ease2024}. The presence of conditionals suggests multiple flows inside a single step or uncertainty about whether the step applies.

\noindent\textbf{Why it is harmful.} Conditionals introduce divergence and can reduce repeatability. Testers may not know whether they should execute alternative actions, skip steps or stop the test. This can lead to incomplete coverage or inconsistent execution across environments and testers~\cite{Hauptmann2013,SoaresAORGSMSFB23}.

\noindent\textbf{Examples.} The following step creates two possible states and does not clarify how the test should proceed in each case.
\begin{stepexample}
<dt>Press the 'Super Windows key', or click the Ubuntu logo in the upper left hand unity panel and search 'Sound Recorder'</dt>
\end{stepexample}

\noindent In the following example, the step depends on a prior condition: the software must have been successfully installed before the tester can log out and verify the language selection options available on the login screen.
\begin{stepexample}
<dt>When they have been successfully installed, log out from the session, and select alternating languages at the login screen, next to the right of the session selector (Xubuntu Session)</dt>
\end{stepexample}

\subsection{Eager Action}
\label{s:eager-action}

\noindent\textbf{Definition.} Eager Action occurs when a single step groups multiple distinct actions that should be separated~\cite{Hauptmann2013,Aranda-ease2024}. In manual tests, this often appears as multiple imperative verbs connected by ``and'', ``then'', commas, or multiple sentences within a single step.

\noindent\textbf{Why it is harmful.} Bundling multiple actions into one step reduces granularity and makes debugging difficult. If a later verification fails, the tester may not know which action caused the failure. It also increases cognitive load and encourages skipping or partially executing instructions~\cite{SoaresAORGSMSFB23,emse-2025}.

\noindent\textbf{Examples.} The following step groups several UI interactions into a single instruction, which makes it harder to isolate failures or confirm that each sub-action was completed correctly.
\begin{stepexample}
<dt>Open the dash and launch rhythmbox by pressing the super key, and then entering 'rhythmbox'</dt>
\end{stepexample}

\noindent In the next example, two sentences still form one step and bundle multiple operations that would be clearer as separate steps.
\begin{stepexample}
<dt>Select a plugin and configure it by doing click on "Configure"</dt>
\end{stepexample}

\subsection{Misplaced Action}
\label{s:misplaced-action}

\noindent\textbf{Definition.} Misplaced Action occurs when an action instruction is written in the verification field, that is, in a step whose role is to state expected outcomes~\cite{SoaresAORGSMSFB23,Aranda-ease2024}. The key signal is imperative action language inside a verification context.

\noindent\textbf{Why it is harmful.} This smell blurs the separation of responsibilities. A verification step should describe what to observe, not what to do. Role confusion increases the likelihood of missing checks, repeating actions, or misreporting failures~\cite{emse-2025}.

\noindent\textbf{Examples.} The following text is an action instruction appearing in a verification step. In this case, the test mixes procedure and oracle information in the wrong field.
\begin{stepexample}
<dd>Open some windows (e.g. terminal)</dd>
\end{stepexample}

\subsection{Misplaced Precondition}
\label{s:misplaced-precondition}

\noindent\textbf{Definition.} Misplaced Precondition occurs when a prerequisite or required state is written as an action or verification step, instead of being stated explicitly as a precondition~\cite{SoaresAORGSMSFB23,Aranda-ease2024}. Common patterns include state descriptions such as the user is logged in, make sure, assuming, given that, or ensure that followed by a state rather than an interaction.

\noindent\textbf{Why it is harmful.} 
Preconditions define when a test is applicable. When they are embedded as ordinary steps, testers may interpret a missing prerequisite as a product failure, which can lead to false bug reports and wasted effort during triage~\cite{Hauptmann2013,emse-2025}.

\noindent\textbf{Examples.} The following sentence describes a precondition state appearing as an action step. In this case, it should be extracted into an explicit precondition.
\begin{stepexample}
<dt>Please ensure the Nautilus window is running and the Nautilus window is focused</dt>
\end{stepexample}

\noindent The next example is a state assertion rather than an action or a verification derived from an action, which indicates that it is a precondition expressed in the wrong location.
\begin{stepexample}
<dt>Ensure that Network Manager is running and that no networks are currently connected</dt>
\end{stepexample}

\subsection{Misplaced Verification}
\label{s:misplaced-verification}

\noindent\textbf{Definition.} Misplaced Verification occurs when a verification or expected outcome is written in the action field~\cite{SoaresAORGSMSFB23,Aranda-ease2024}. Typical cues include verbs such as verify, check, confirm, ensure, or outcome statements with should, must, is displayed, appears, or no error.

\noindent\textbf{Why it is harmful.} Mixing checks into action steps can lead testers to treat verification as optional or to perform checks at the wrong time. It also complicates automated processing because the same field contains both procedure and oracle information~\cite{Hauptmann2013,emse-2025}.

\noindent\textbf{Examples.} The following step is phrased as a check and appears in the action field. In this case, it is a verification written in the wrong location.
\begin{stepexample}
<dt>Run 'apt-get update' and verify that an ec2 mirror is used</dt>
\end{stepexample}

\noindent The next example states an expected outcome rather than an interaction and is written in the action field. Therefore, it should be expressed as a verification step.
\begin{stepexample}
<dt>Once the machine boots up, it should boot into the frontend</dt>
\end{stepexample}

\subsection{Unverified Action}
\label{s:unverified-action}

\noindent\textbf{Definition.} Unverified Action occurs when an action step has no corresponding verification that states what the tester should observe after performing that action~\cite{Hauptmann2013,Aranda-ease2024}. In structured manual tests, this is often identified by an action that is not followed by any verification step before the next action begins.

\noindent\textbf{Why it is harmful.} Without an explicit oracle, the tester cannot determine whether the action succeeded or whether the system behaved correctly. This invites implicit assumptions and increases variability across testers, which reduces test reliability and weakens the evidential value of manual execution~\cite{emse-2025}.

\noindent\textbf{Examples.} The following instruction describes an interaction but provides no indication of what should be observed next. Without a subsequent verification step, the action is effectively unverified.
\begin{stepexample}
<dt>Click the Search Icon in the top</dt>
<dt>Input 'test'</dt>
    <dd>the 'test' folder was shown?</dd>
\end{stepexample}

\noindent In the next example, the test instructs the tester to open the application. Without an immediate verification describing the expected screen or status, the notion of success remains underspecified.
\begin{stepexample}
<dt>Open the file manager</dt>
\end{stepexample}

Table~\ref{tab:catalogued-smells} summarizes the seven smells targeted by our prompt and used in our evaluation. The definitions are concise by design, and the preceding subsections provide rationale and concrete examples to support consistent interpretation.

\begin{table}[t]
\centering
\caption{Catalogued test smells considered in this article.}
\label{tab:catalogued-smells}
\begin{tabular}{p{0.30\linewidth} p{0.64\linewidth}}
\hline
\textbf{Test Smell} & \textbf{Brief definition} \\
\hline
Ambiguous Test & Test step is vague or underspecified, leaving room for interpretation. \\
Conditional Test & Step includes conditional logic expressed in natural language. \\
Eager Action & Single step groups multiple distinct actions that should be separated. \\
Misplaced Action & Action instruction written as a verification step. \\
Misplaced Precondition & Prerequisite or required state written as an action or verification step. \\
Misplaced Verification & Verification or expected outcome written as an action step. \\
Unverified Action & Action step without a corresponding verification describing expected outcome. \\
\hline
\end{tabular}
\end{table}

\subsection{Practical Relevance}
\label{practical-relevance}

Evidence from the Ubuntu manual-tests repository suggests that the issues captured by our catalogue of test smells arise in practice and are actively addressed during routine test maintenance.
For instance, in a bug-fix commit (Fix bug \#2012346), maintainers refined a manual test case by clarifying the instructions, splitting a previously coarse step into multiple, more focused test steps, and adding an explicit expected outcome (i.e., a verification that UI elements appear in the selected language).%
\footnote{\url{https://git.launchpad.net/ubuntu-manual-tests/commit/?id=def6df4d810dedeea53fc3a63796aa2c717b33fb}}. 
This revision improves precision and reduces the risk of underspecified or unverified actions.
At the same time, the patch also illustrates how smells can be inadvertently introduced during editing: the following newly added step expresses a check but is recorded as an \emph{action} step (i.e., it appears in a \texttt{<dt>} field), which corresponds to our \textit{Misplaced Verification} category.
\begin{stepexample}
<dt>[...] and verify that an ec2 mirror is used</dt>
\end{stepexample}
Overall, such changes likely reflect manual review and iterative refinement by practitioners, yet they also highlight the lack of lightweight support for reasoning about the quality of natural language test cases.
Our goal is to automate the detection of these issues, enabling test authors and reviewers to identify ambiguous, conditional, misplaced, eager, or unverified steps earlier and with lower effort.

\section{Methodology}
\label{sec:methodology}

This section describes the methodology adopted to evaluate the ability of foundation models to detect natural language test smells in Ubuntu-style manual test cases. We define our study goals and research question (RQ) following the Goal Question Metric (GQM) paradigm in Section~\ref{sec:gqm}. We then present the dataset and annotation procedure in Section~\ref{sec:methodology-dataset}, describe the prompting strategy used to operationalize the test smell definitions in Section~\ref{s:methodology-prompt}, and detail the evaluated models and evaluation procedure in Sections~\ref{sec:methodology-model}~and~\ref{sec:methodology-eval}, respectively. 

\subsection{Goal Question Metric}
\label{sec:gqm}

We structure our study using the Goal Question Metric (GQM) paradigm~\cite{Basili1994}. Our goal is to evaluate whether a contemporary LLM can accurately identify test smells in manual test descriptions while providing explanations that are useful to practitioners.

\noindent\textbf{Goal.} Analyze \gemini{} for the purpose of evaluating its effectiveness in detecting natural language test smells in Ubuntu-style manual test cases with respect to detection performance and explanation usefulness from the viewpoint of software testers and practitioners in the context of real-world manual testing artifacts.

\noindent Based on this goal, we derive the following RQ:
\begin{itemize}
\item[\textbf{RQ$_{1}$}] To what extent can \gemini{} detect natural language test smells in Ubuntu-style manual test cases?
\end{itemize}

\noindent\textbf{Metrics.} To answer RQ$_{1}$, we report: (i) aggregated statistics on smell prevalence, including the average number of detected smells per test-case step, to characterize their frequency in practice; and (ii) detection performance using standard classification metrics, namely precision, recall, and F$_1$ score, computed per smell type and overall. Our ground truth labels were obtained through manual annotation: from the \numStepsTotal{} test steps in our dataset, one author annotated a subset of \numStepsManual{} test steps from \numTestsManual{} test cases following the definitions detailed in Section~\ref{sec:test-smells}. Additionally, we analyze explanation usefulness by verifying whether model-generated explanations identify concrete textual triggers aligned with the smell definitions

\subsection{Dataset}
\label{sec:methodology-dataset}

Our dataset is drawn from the Ubuntu Manual Tests repository~\cite{ubuntu_manual_tests}, which contains community-maintained manual test cases used by the Ubuntu QA (Quality Assurance) team to support release and regression activities, including ISO image testing and broader system validation. Ubuntu is a widely used Linux distribution with a large global user base, and its QA infrastructure relies on these manual test specifications to coordinate execution and record results through instances of the QATracker platform~\cite{Hauptmann2013,SoaresAORGSMSFB23}. In this study, we analyze \numTestCases{} manual test cases from this repository.

\subsection{Prompting Strategy}
\label{s:methodology-prompt}

We employ Meta Prompting~\cite{hou_metaprompting,prompt-techniques} to guide the model in generating and refining our task prompt. We used \gpt{} in January 2026 to improve the prompt used before~\cite{keila-sbes-2025}. Then, we used \gemini{} to improve the prompt generated by \gpt{}. The resulting prompt combines a domain-specific persona (QA engineer and test smell detector), explicit instructions and constraints, and a small set of few-shot examples to calibrate the desired behavior and output format. This process results in the final prompt, which we present next.

\begin{tcolorbox}[
  breakable,
  colback=gray!5,
  colframe=black!60,
  boxrule=0.5pt,
  arc=2pt,
  left=6pt,right=6pt,top=6pt,bottom=6pt
]
\footnotesize
You are an expert QA Engineer and Natural Language Test Smell Detector. \\
\\
\noindent TASK \\
\noindent Analyze ONE complete manual test case (Ubuntu manual test style) and detect Natural Language Test Smells for EACH step, based ONLY on the step text and the step location derived from the markup: \\
\noindent - \texttt{<dt> ... </dt>} \ \ \ $\Rightarrow$ \texttt{where = "action"} \\
\noindent - \texttt{<dd> ... </dd>} \ \ \ $\Rightarrow$ \texttt{where = "verification"} \\
\\
\noindent IMPORTANT INPUT SHAPE \\
\noindent You will receive the full test case as a single text block that may contain: \\
\noindent - A title or intro line (plain text) \\
\noindent - An HTML like \texttt{<dl>} list with multiple \texttt{<dt>} and \texttt{<dd>} elements \\
\noindent - Other HTML tags (e.g., \texttt{<strong>}, \texttt{<a>}, etc.) that are NOT steps \\
\\
\noindent You must: \\
\noindent 1) Extract ONLY the text inside each \texttt{<dt>} and \texttt{<dd>} (strip tags inside them, keep the visible text). \\
\noindent 2) Treat each \texttt{<dt>} or \texttt{<dd>} as ONE step item, even if it contains multiple sentences. \\
\noindent 3) Ignore any content outside \texttt{<dl> ... </dl>} for smell detection (including instructions like \texttt{If all actions produce...}). \\
\\
\noindent OUTPUT (STRICT) \\
\noindent Return ONLY a valid JSON object (no markdown, no extra text) with EXACTLY these fields: \\
\noindent \texttt{\{} \\
\noindent \ \ \texttt{"steps": [} \\
\noindent \ \ \ \ \ \texttt{\{} \\
\noindent \ \ \ \ \ \ \ \texttt{"id": 1,} \\
\noindent \ \ \ \ \ \ \ \texttt{"where": "action",} \\
\noindent \ \ \ \ \ \ \ \texttt{"text": "Step text as plain text",} \\
\noindent \ \ \ \ \ \ \ \texttt{"smells": ["Smell Name 1", "Smell Name 2"],} \\
\noindent \ \ \ \ \ \ \ \texttt{"explanation": "Concise reasoning citing the exact words or patterns that triggered the smell(s)."} \\
\noindent \ \ \ \ \ \texttt{\}} \\
\noindent \ \ \texttt{]} \\
\noindent \texttt{\}} \\
\\
\noindent RULES FOR THE JSON OUTPUT \\
\noindent - \texttt{"id"} must start at 1 and increment by 1 following the order in the \texttt{<dl>}. \\
\noindent - If no smells apply to a step: \\
\noindent \ \ \ \texttt{"smells": []} \\
\noindent \ \ \ \texttt{"explanation": "No clear smell indicators in the step given its field."} \\
\noindent - Do NOT add any other JSON fields (no title, no summary, no counts). \\
\\
\noindent GENERAL RULES (STRICT) \\
\noindent - Be conservative: label a smell only when there is clear evidence in the step text. \\
\noindent - You may assign multiple smells to the same step. \\
\noindent - Always use the smell names EXACTLY as listed below. \\
\noindent - Explanations must be short (one or two reasons) and MUST cite exact triggers from the step text \\
\noindent \ \ (e.g., \texttt{contains "if"}, \texttt{contains "Verify"}, \texttt{contains "any"}, \texttt{multiple imperative verbs: "open", "click"}). \\
\noindent - If \texttt{where="action"} and the step mixes actions and verification language, label BOTH \texttt{"Eager Action"} and \texttt{"Misplaced Verification"}. \\
\noindent - Do NOT infer context outside the step text. \\
\noindent - Do NOT use surrounding steps to decide a smell for the current step, \\
\noindent \ \ EXCEPT for \texttt{"Unverified Action"}, which requires checking whether a \texttt{<dt>} action step \\
\noindent \ \ has at least one corresponding \texttt{<dd>} verification step associated to it by structure. \\
\\
\noindent SPECIAL CASE FOR \texttt{"Unverified Action"} \\
\noindent For \texttt{"Unverified Action"} only, you may use the \texttt{<dl>} order to check whether a \texttt{<dt>} action step \\
\noindent is followed by a corresponding \texttt{<dd>} verification step (before the next \texttt{<dt>}). \\
\noindent For all other smells, do not use surrounding steps. \\
\\
\noindent SMELLS AND RULES (STRICT) \\
\\
\noindent 1) \texttt{"Ambiguous Test"} \\
\noindent Definition: Vague, subjective, or underspecified step; pass or fail is not objectively verifiable. \\
\noindent Mark if ANY trigger applies: \\
\noindent A) Verb + indefinite determiner or vague quantity targeting an open-ended object: \\
\noindent \ \ - Patterns like: \texttt{"Open any ..."}, \texttt{"Select a ..."}, \texttt{"Choose some ..."}, \texttt{"Pick several ..."} \\
\noindent \ \ - Indefinite determiners or quantities: \texttt{"any"}, \texttt{"a"}, \texttt{"an"}, \texttt{"some"}, \texttt{"a few"}, \texttt{"several"}, \texttt{"various"}, \texttt{"a lot"} \\
\noindent \ \ - IMPORTANT: Only mark this when the object is not uniquely identified \\
\noindent \ \ \ \ (no clear label or identifier like quoted UI text, exact value, ID, or precise menu path). \\
\noindent B) Indefinite pronouns or unclear referent: \\
\noindent \ \ - \texttt{"something"}, \texttt{"someone"}, \texttt{"somewhere"}, \texttt{"anything"}, \texttt{"anyone"}, \texttt{"anywhere"}, \texttt{"everything"} \\
\noindent \ \ - \texttt{"it/this/that"} ONLY if the referent is unclear inside the same step text. \\
\noindent C) Subjective adverbs or adjectives or phrases: \\
\noindent \ \ - \texttt{"as expected"}, \texttt{"properly"}, \texttt{"correctly"}, \texttt{"user-friendly"}, \texttt{"randomly"}, \texttt{"normally"}, \texttt{"adequately"}, \texttt{"quickly/fast"} \\
\noindent D) Comparative or superlative language without objective metric: \\
\noindent \ \ - \texttt{"better"}, \texttt{"worse"}, \texttt{"faster"}, \texttt{"more stable"}, \texttt{"best"}, \texttt{"most"}, \texttt{"least"} \\
\\
\noindent 2) \texttt{"Conditional Test"} \\
\noindent Definition: Introduces branching logic or depends on a condition, implying multiple flows. \\
\noindent Mark if the step contains conditional markers: \\
\noindent \ \ \texttt{"if"}, \texttt{"when"}, \texttt{"unless"}, \texttt{"in case"}, \texttt{"depending on"}, \texttt{"whether"}, \texttt{"as long as"}, \texttt{"otherwise"}, \texttt{"only if"} \\
\\
\noindent 3) \texttt{"Eager Action"} \\
\noindent Definition: A single step groups multiple distinct actions that should be separate. \\
\noindent Mark if there are two or more distinct operations, signaled by: \\
\noindent \ \ - multiple imperative verbs, and or connectors: \texttt{"and"}, \texttt{"then"}, commas, \texttt{"after"}, \texttt{"followed by"} \\
\noindent \ \ - multiple sentences within the same step that contain multiple operations \\
\noindent Examples of action verbs: \texttt{Open, Click, Select, Type, Navigate, Install, Run, Adjust, Press, Tap, Choose, Enter} \\
\\
\noindent 4) \texttt{"Misplaced Action"} \\
\noindent Definition: Action instructions written in the verification field. \\
\noindent Mark ONLY IF: \\
\noindent \ \ - \texttt{where == "verification"} \\
\noindent \ \ - AND the step contains imperative action verbs (e.g., \texttt{"Click"}, \texttt{"Open"}, \texttt{"Type"}, \texttt{"Select"}, \texttt{"Adjust"}, \texttt{"Install"}, \texttt{"Navigate"}) \\
\\
\noindent 5) \texttt{"Misplaced Precondition"} \\
\noindent Definition: Prerequisite or setup or state written in action or verification instead of precondition. \\
\noindent Mark ONLY IF: \\
\noindent \ \ - \texttt{where} in \texttt{["action","verification"]} \\
\noindent \ \ - AND the step is primarily a prerequisite or state (not an action), e.g.: \\
\noindent \ \ \ \ - SUT state pattern: \texttt{"X is/are/was/were/has/have ..."} + adjective or past participle (e.g., \texttt{"The user is logged in."}) \\
\noindent \ \ \ \ - explicit cues: \texttt{"Precondition:"}, \texttt{"Prerequisite:"}, \texttt{"Make sure"}, \texttt{"Assuming"}, \texttt{"Given that"}, \texttt{"Ensure"} (state) \\
\\
\noindent 6) \texttt{"Misplaced Verification"} \\
\noindent Definition: Verification or checking written in the action field. \\
\noindent Mark ONLY IF: \\
\noindent \ \ - \texttt{where == "action"} \\
\noindent \ \ - AND the step includes verification verbs or outcome assertions: \\
\noindent \ \ \ \ - verification verbs: \texttt{"Verify"}, \texttt{"Check"}, \texttt{"Ensure"}, \texttt{"Confirm"}, \texttt{"Validate"}, \texttt{"Observe"} \\
\noindent \ \ \ \ - outcome language: \texttt{"should"}, \texttt{"must"}, \texttt{"is displayed"}, \texttt{"appears"}, \texttt{"changes"}, \texttt{"is updated"}, \texttt{"is saved"}, \texttt{"no error"}, \texttt{"successfully"} \\
\\
\noindent 7) \texttt{"Unverified Action"} \\
\noindent Definition: Action steps that miss corresponding verification steps. \\
\noindent Problem: Absent verification steps negatively affect test execution and correctness since there is no instruction on how the system should behave, leaving room for the testers' interpretation. \\
\noindent Identification (STRICT): Mark ONLY IF: \\
\noindent \ \ - \texttt{where == "action"} \\
\noindent \ \ - AND the step is an action instruction (contains imperative action verb like \texttt{"Open"}, \texttt{"Click"}, \texttt{"Select"}, \texttt{"Type"}, \texttt{"Navigate"}, \texttt{"Install"}, \texttt{"Run"}, \texttt{"Press"}, etc.) \\
\noindent \ \ - AND the step does NOT contain verification or outcome language by itself \\
\noindent \ \ \ (e.g., \texttt{"Verify"}, \texttt{"Check"}, \texttt{"Ensure"}, \texttt{"Confirm"}, \texttt{"should"}, \texttt{"is displayed"}, \texttt{"appears"}, \texttt{"no error"}, \texttt{"successfully"}) \\
\noindent \ \ - AND there is NO corresponding verification step structurally linked to it: \\
\noindent \ \ \ \ - i.e., there is no \texttt{<dd>} step immediately following this \texttt{<dt>} before the next \texttt{<dt>}, in the extracted \texttt{<dl>} order. \\
\noindent Notes to reduce false positives: \\
\noindent \ \ - If the action step itself contains verification or outcome language, do NOT mark \texttt{"Unverified Action"}. \\
\noindent \ \ \ \ Instead, if \texttt{where == "action"}, mark \texttt{"Misplaced Verification"} and possibly \texttt{"Eager Action"} per the rules above. \\
\noindent \ \ - Do not search for a verification elsewhere in the test case. Only consider the immediately following \texttt{<dd>} as the correspondence rule. \\
\\
\noindent FEW-SHOT EXAMPLES (FULL TEST CASE STYLE) \\
\\
\noindent Input test case: \\
\noindent \texttt{<dl>} \\
\noindent \ \ \texttt{<dt>Open the browser and navigate to Google.</dt>} \\
\noindent \ \ \texttt{<dt>Select the Settings menu and click Advanced.</dt>} \\
\noindent \ \ \texttt{<dd>The Advanced settings are displayed.</dd>} \\
\noindent \texttt{</dl>} \\
\\
\noindent Output: \\
\noindent \texttt{\{} \\
\noindent \ \ \texttt{"steps": [} \\
\noindent \ \ \ \ \ \texttt{\{ "id": 1, "where": "action", "text": "Open the browser and navigate to Google.", "smells": ["Eager Action","Unverified Action"], "explanation": "Multiple imperative verbs: 'Open' and 'navigate' connected by 'and'. No immediate <dd> verification follows this <dt>." \},} \\
\noindent \ \ \ \ \ \texttt{\{ "id": 2, "where": "action", "text": "Select the Settings menu and click Advanced.", "smells": ["Eager Action"], "explanation": "Multiple imperative verbs: 'Select' and 'click' connected by 'and'. An immediate <dd> verification follows this <dt>." \},} \\
\noindent \ \ \ \ \ \texttt{\{ "id": 3, "where": "verification", "text": "The Advanced settings are displayed.", "smells": [], "explanation": "No clear smell indicators in the step given its field." \}} \\
\noindent \ \ \texttt{]} \\
\noindent \texttt{\}} \\
\\
\noindent NOW ANALYZE THIS INPUT TEST CASE \\
\noindent \texttt{<test\_case>} \\
\noindent \texttt{\{test\}} \\
\noindent \texttt{</test\_case>}
\end{tcolorbox}

For clarity, the phrase ``not an action'' is intended to exclude not only procedural actions but also verification statements and outcome assertions. Verification-oriented text is captured separately by the Misplaced Verification smell, whereas Misplaced Precondition is reserved for prerequisite states and setup conditions that should appear in the precondition section of the test case.

In the prompt, \texttt{test} represents the Ubuntu test case. For example, \texttt{test} corresponds to the test case in Listing~\ref{lst:ubuntu-dl}. To guide \gemini{} in detecting test smells, we designed a structured prompt that (i) embeds the definitions of the seven smells considered in this study, grounded in the definitions used in prior work~\cite{Hauptmann2013,SoaresAORGSMSFB23,Aranda-ease2024}, and (ii) explicitly encodes the Ubuntu manual test representation. In particular, the prompt explains that Ubuntu test cases are provided as an HTML-like \texttt{<dl>} block, where each \texttt{<dt>} element corresponds to an \emph{action} step (what the tester does) and each \texttt{<dd>} element corresponds to a \emph{verification} step (what the tester checks). The model is instructed to extract only the text inside \texttt{<dt>} and \texttt{<dd>}, treat each extracted element as a single step item, and ignore any content outside the \texttt{<dl>} block. The prompt enforces a strict JSON schema with one entry per step, including the step role, the predicted smell labels, and a short explanation that cites concrete textual triggers (e.g., the presence of conditional markers, vague quantifiers, or multiple imperative verbs). Finally, the prompt includes a special rule for \emph{Unverified Action}, allowing the model to use only the local \texttt{<dl>} order to determine whether an action step is followed by a corresponding verification step, while requiring all other smells to be decided solely from the current step text.

\subsection{Model}
\label{sec:methodology-model}

We selected \gemini{} (Google) because it is widely used in practice and is consistently ranked among the most competitive models on LLM Arena~\cite{llm-arena}. We accessed the model through the official API and used the default inference settings provided by the API client. All experiments were conducted between January and February 2026.

\subsection{Evaluation Procedure}
\label{sec:methodology-eval}
For each test case, we run \gemini{} with our prompting strategy and obtain, for every test step, (i) a predicted set of smell labels and (ii) a short, evidence-grounded explanation. We evaluate a subset of these predictions by comparing the model-assigned labels against a manually established reference. Concretely, one author labeled the selected steps according to the smell definitions in Section~\ref{sec:test-smells}; we treat these annotations as ground truth to compute precision, recall, and F$_1$ score. We consider a prediction correct if the model assigns \emph{exactly} the same set of smell labels as the manual annotation for the corresponding step. In addition to label agreement, we qualitatively inspect explanations to verify whether they cite explicit textual cues (e.g., keywords, patterns, or phrasing) that are consistent with the corresponding smell definitions.

Given the cost of manual labeling, we manually inspected $n=\amostraManual{}$ test steps drawn from 15 test cases, out of a population of $N=\numStepsTotal{}$ steps. Under a finite-population assumption and worst-case proportion ($p=0.5$), this sample size yields an approximate 95\% margin of error of about $\pm 10$ percentage points when estimating step-level proportions. To improve coverage under a limited labeling budget, we adopted a stratified sampling strategy that includes both smelly and non-smelly steps and prioritizes the most frequent smell categories. We therefore treat this manual assessment as an initial, cost-effective validation, and we discuss larger-scale annotation as future work.

\section{Results}
\label{sec:results}

This section presents the results of our empirical evaluation. Section~\ref{sec:rq1} characterizes the prevalence of test smells across the analyzed Ubuntu test cases and reports the smells identified by \gemini{}. Section~\ref{sec:manual-analysis} compares the model predictions against a manually annotated baseline to assess detection performance. We then compare \gemini{} with \gpt{} in Section~\ref{sec:gpt}, discuss the relationship between our findings and previous results obtained with SLMs in Section~\ref{sec:slms}, and compare our approach with the rule-based Manual Test Sensei tool in Section~\ref{sec:related-sensei}. Finally, Section~\ref{sec:threats} discusses the main threats to validity of our study.

\subsection{RQ$_{1}$. \gemini{}}
\label{sec:rq1}

We report the results of manual test smell detection using \gemini{} as the underlying model. The analyzed corpus contains \numTestCases{} manual test cases with \numStepsTotal{} test steps in total. Across all steps, \gemini{} flagged 2,293 smell instances (i.e., a step may contain multiple smells). Out of the \numStepsTotal{} steps, 1,533 steps (58.56\%) contain at least one smell, while 1,085 steps (41.44\%) contain no smells. This indicates that smell occurrences are common at the step level: more than half of the steps are affected by at least one natural language issue.

We quantify density in two complementary ways.
First, the overall smell density is 0.876 smells per step.
Second, considering only steps that exhibit at least one smell, \gemini{} reports an average of 1.496 smells per smelly step, indicating that multi-smell steps are common.
Smell counts also vary substantially across test cases.
The minimum is one smell, observed in \texttt{packages/1423\_LibreOffice/libreoffice/lbc-004}.\footnote{\href{https://github.com/easy-software-ufal/manual-test-sensei/blob/main/testcases/packages/1423_LibreOffice}{Manual test case \#1423 (packages)}}
This test case contains only two steps, and exactly one is flagged with Eager Action.
\begin{stepexample}
<dt>Open the dash and launch LibreOffice Writer by pressing the super key, and then entering 'writer'</dt>
\end{stepexample}
\noindent \gemini{} produces the following output:
\begin{geminioutput}
\textbf{Smells:} Eager Action \\
\textbf{Explanation:} Multiple imperative verbs and actions: `Open', `launch', `pressing', `entering' connected by `and', `and then'. 
\end{geminioutput}

As another example, consider the test case
\texttt{image/1461\_MAAS (Installer)/maas/maas-001}.\footnote{\href{https://github.com/easy-software-ufal/manual-test-sensei/blob/main/testcases/image/1461_MAAS\%20(Installer)}{Manual test case \#1461 (image)}} This test case includes 13 test steps, and each step is annotated with up to three test smells. For instance, consider the second action step:
\begin{stepexample}
<dt>Please ensure you have full control over the network</dt>
\end{stepexample}

\noindent \gemini{} returns the following prediction:

\begin{geminioutput}
\textbf{Smells:} Ambiguous Test; Misplaced Precondition; Misplaced Verification\\
\textbf{Explanation:} Contains subjective phrase `full control'. Step describes a prerequisite state (`Please ensure you have...'). Contains verification verb `ensure' in the action field.
\end{geminioutput}

In this case, the test case does not provide an explicit precondition field. As a result, \gemini{} cannot disambiguate whether the sentence should be modeled as (i) a precondition that was placed in an action step, or (ii) a verification statement that should appear as a \texttt{<dd>} but was mistakenly written as a \texttt{<dt>}. Consequently, \gemini{} reports both Misplaced Precondition and Misplaced Verification. It also flags Ambiguous Test, since the phrase ``full control over the network'' is underspecified and does not define an objective, verifiable condition.

At the step level, the maximum number of smells observed in a single test step is 4. At the test-case level, the maximum is 101 test smells, observed in
\texttt{1460\_iSCSI Authenticated Root Installation}.\footnote{\href{https://github.com/easy-software-ufal/manual-test-sensei/blob/main/testcases/image/1460_iSCSI\%20Authenticated\%20Root\%20Installation}
{Manual test case \#1460 (image)}}
This test case includes 51 test steps, and each step is annotated with one to three test smells. 
For instance, consider the following action step:

\begin{stepexample}
<dt>On Keyboard Layout select No and press Enter</dt>
\end{stepexample}

\noindent \gemini{} produces the following output:

\begin{geminioutput}
\textbf{Smells:} Eager Action; Unverified Action\\
\textbf{Explanation:} Multiple imperative verbs `select' and `press' connected by `and'. No immediate <dd> verification follows this <dt>.
\end{geminioutput}

\begin{figure}[t]
\centering
\begin{tikzpicture}
\begin{axis}[
    boxplot/draw direction=y,
    width=\linewidth,
    height=6.5cm,
    ymajorgrids,
    grid style={dashed},
    ymin=0, ymax=105,
    xtick={1,2,3,4},
    xticklabels={
        test steps,
        steps with smells,
        steps without smells,
        test smells 
    },
    xticklabel style={rotate=25, anchor=east},
    ylabel={Count per test case},
]

\addplot+[
    draw=blue,
    boxplot prepared={
      lower whisker=2,
      lower quartile=10,
      median=23,
      upper quartile=36.5,
      upper whisker=73
    }
] coordinates {};
\addplot+[
    only marks,
    mark=*,
    mark size=1.2pt,
    draw=blue,
    fill=blue,
    forget plot
] coordinates {(1,84) (1,85)};

\addplot+[
    draw=red,
    boxplot prepared={
      lower whisker=1,
      lower quartile=6,
      median=11.5,
      upper quartile=22.25,
      upper whisker=37
    }
] coordinates {};
\addplot+[
    only marks,
    mark=*,
    mark size=1.2pt,
    draw=red,
    fill=red,
    forget plot
] coordinates {(2,50) (2,51) (2,59) (2,69)};

\addplot+[
    draw=brown,
    boxplot prepared={
      lower whisker=0,
      lower quartile=1.75,
      median=5,
      upper quartile=15.5,
      upper whisker=35
    }
] coordinates {};
\addplot+[
    only marks,
    mark=*,
    mark size=1.2pt,
    draw=brown,
    fill=brown,
    forget plot
] coordinates {(3,39) (3,44) (3,45) (3,46) (3,57)};

\addplot+[
    draw=black,
    boxplot prepared={
      lower whisker=1,
      lower quartile=9,
      median=19.5,
      upper quartile=33.5,
      upper whisker=66
    }
] coordinates {};
\addplot+[
    only marks,
    mark=*,
    mark size=1.2pt,
    draw=black,
    fill=black,
    forget plot
] coordinates {(4,74) (4,99) (4,101)};

\end{axis}
\end{tikzpicture}
\caption{Aggregated absolute metrics per test case.}
\label{fig:boxplots_aggregated_absolute}
\end{figure}
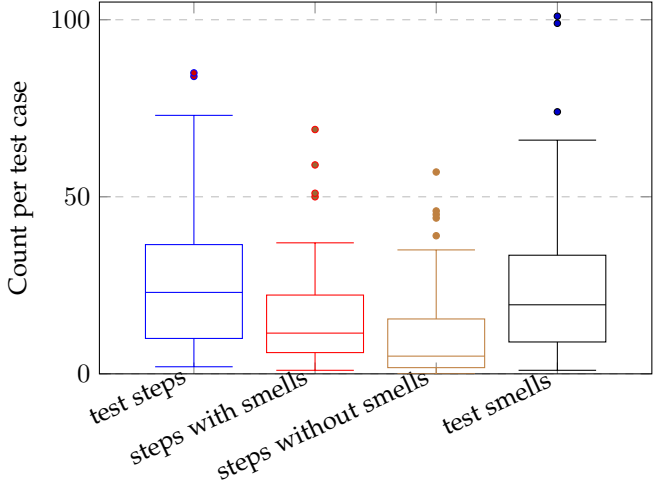

In Figure~\ref{fig:boxplots_aggregated_absolute}, the number of test steps per test case varies moderately (median $=23$), with a few longer outliers reaching 84--85 steps; similarly, total test smell instances per test case have median $=19.5$ and show a heavy tail with outliers up to 101.

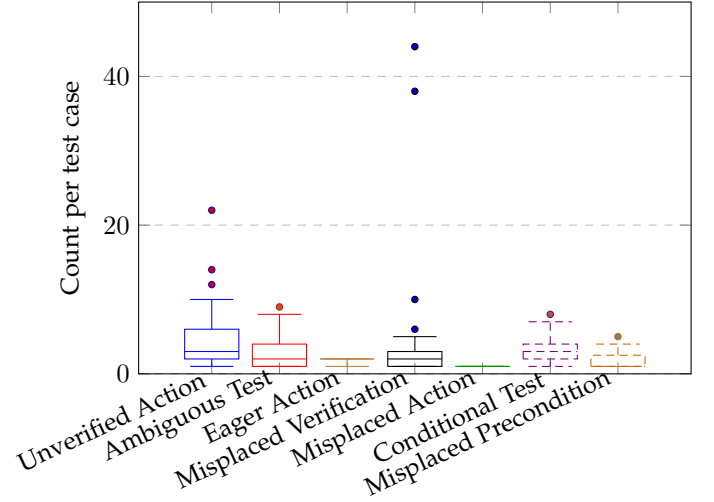
\begin{figure}[t]
\centering
\begin{tikzpicture}
\begin{axis}[
    boxplot/draw direction=y,
    width=\linewidth,
    height=6.5cm,
    ymajorgrids,
    grid style={dashed},
    ymin=0, ymax=50,
    xtick={1,2,3,4,5,6,7},
    xticklabels={
        Unverified Action ,
        Ambiguous Test ,
        Eager Action ,
        Misplaced Verification ,
        Misplaced Action ,
        Conditional Test ,
        Misplaced Precondition 
    },
    xticklabel style={rotate=25, anchor=east},
    ylabel={Count per test case},
]

\addplot+[
    draw=blue,
    boxplot prepared={lower whisker=1, lower quartile=2, median=3, upper quartile=6, upper whisker=10}
] coordinates {};
\addplot+[only marks, mark=*, mark size=1.2pt, draw=blue, fill=blue, forget plot]
  coordinates {(1,12) (1,14) (1,22)};

\addplot+[
    draw=red,
    boxplot prepared={lower whisker=1, lower quartile=1, median=2, upper quartile=4, upper whisker=8}
] coordinates {};
\addplot+[only marks, mark=*, mark size=1.2pt, draw=red, fill=red, forget plot]
  coordinates {(2,9)};

\addplot+[
    draw=brown,
    boxplot prepared={lower whisker=1, lower quartile=2, median=2, upper quartile=2, upper whisker=2}
] coordinates {};

\addplot+[
    draw=black,
    boxplot prepared={lower whisker=1, lower quartile=1, median=2, upper quartile=3, upper whisker=5}
] coordinates {};
\addplot+[only marks, mark=*, mark size=1.2pt, draw=black, fill=black, forget plot]
  coordinates {(4,6) (4,10) (4,38) (4,44)}; 

\addplot+[
    draw=green!60!black,
    boxplot prepared={lower whisker=1, lower quartile=1, median=1, upper quartile=1, upper whisker=1}
] coordinates {};

\addplot+[
    draw=violet,
    boxplot prepared={lower whisker=1, lower quartile=2, median=3, upper quartile=4, upper whisker=7}
] coordinates {};
\addplot+[only marks, mark=*, mark size=1.2pt, draw=violet, fill=violet, forget plot]
  coordinates {(6,8)};

\addplot+[
    draw=orange!80!black,
    boxplot prepared={lower whisker=1, lower quartile=1, median=1, upper quartile=2.5, upper whisker=4}
] coordinates {};
\addplot+[only marks, mark=*, mark size=1.2pt, draw=orange!80!black, fill=orange!80!black, forget plot]
  coordinates {(7,5)};

\end{axis}
\end{tikzpicture}
\caption{Absolute test smell counts per test case.}
\label{fig:boxplots_smells_absolute}
\end{figure}

Complementarily, Figure~\ref{fig:boxplots_smells_absolute} shows the distribution of absolute counts for each smell type across test cases. The distributions shown in Figure~\ref{fig:boxplots_smells_absolute} indicate that Unverified Action, Ambiguous Test, Eager Action, Misplaced Verification, and Conditional Test are recurrent across the analyzed test cases, whereas Misplaced Action and Misplaced Precondition tend to appear less frequently. The presence of several high-value outliers further indicates that some test cases accumulate substantially more smell instances than others. Together, Figures~\ref{fig:boxplots_aggregated_absolute} and~\ref{fig:boxplots_smells_absolute} suggest that natural language test smells are widespread throughout the corpus, although their distribution is uneven across both test cases and smell categories.

This spread suggests that some test cases are comparatively clean, whereas others concentrate a large number of issues and would likely require substantial rewriting to improve readability and verifiability. At the corpus level, the three most common smells are Unverified Action (580), Eager Action (563), and Ambiguous Test (465). See more details in Table~\ref{tab:gemini-prevalence}. Together, these categories account for the majority of all detected smell instances, indicating that verification clarity, step granularity, and ambiguity are recurrent problems in the analyzed test steps.

Beyond counting how many times each smell occurs at the step level, we also measure \emph{prevalence} at the test-case level: for each smell, we compute the percentage of test cases (out of \numTestCases{}) that contain at least one instance of that smell. Table~\ref{tab:gemini-prevalence} shows that test smells are widespread across the analyzed manual test cases. The most prevalent smells are Eager Action (94\%) and Ambiguous Test (92\%), indicating that most test cases include steps that either bundle multiple actions or rely on vague and subjective wording. Misplaced Verification is also frequent (76\%), suggesting that verification statements are often written in the wrong field (e.g., expressed as actions). Mid-level prevalence smells include Conditional Test (65\%) and Unverified Action (64\%), highlighting that conditional branching and missing/insufficient verification occur in roughly two thirds of the test cases. In contrast, Misplaced Action (26\%) and Misplaced Precondition (23\%) are less common, implying that field misplacement affects a smaller subset of the corpus.

\begin{table}[t]
  \centering
  \caption{Prevalence of each test smell across \numTestCases{} test cases (percentage of test cases with at least one instance) and total occurrences in all steps.}
  \label{tab:gemini-prevalence}
  \small
  \begin{tabular}{lrr}
    \toprule
    \textbf{Test smell} & \textbf{Count} & \textbf{Prevalence (\%)} \\
    \midrule
    Eager Action & 563 & 94\% \\
    Ambiguous Test & 465 & 92\% \\
    Misplaced Verification & 303 & 76\% \\
    Conditional Test & 200 & 65\% \\
    Unverified Action & 580 & 64\% \\
    Misplaced Action & 139 & 26\% \\
    Misplaced Precondition & 43 & 23\% \\
    \bottomrule
  \end{tabular}
\end{table}

\subsection{Manual Analysis}
\label{sec:manual-analysis}

It is important to interpret our results against a baseline. However, since the dataset contains \numStepsTotal{} test steps, manually reviewing all steps would be prohibitively time-consuming. Therefore, the first author manually analyzed \numStepsManual{} steps sampled from \numTestsManual{} test cases and used these annotations as a baseline. Using this baseline, our \gemini{}-based approach achieves \precision{} precision and \recall{} recall, with an F$_1$ score of \fscore{}.

Overall, \gemini{} correctly detected most smell categories. In some cases, the model identifies local cues (e.g., multiple verbs or embedded assertions) but misses the broader pattern of missing verification. For example, consider the test case
\texttt{1482\_Wireless Networking/hwn-001}.\footnote{\href{https://github.com/easy-software-ufal/manual-test-sensei/blob/main/testcases/hardware/1482_Wireless\%20Networking}{Manual test case \#1482 (hardware)}}
This test case includes 11 steps. Consider the following action step:

\begin{stepexample}
<dt>Run <code>route -n</code>. Note the IP address in the Gateway column. (not "0.0.0.0")</dt>
\end{stepexample}

\noindent The annotation produced by \gemini{} is as follows:

\begin{geminioutput}
\textbf{Smells:} Eager Action; Misplaced Verification\\
\textbf{Explanation:} Multiple imperative verbs: `Run' and `Note'. Contains outcome assertion `(not ``0.0.0.0'')'
\end{geminioutput}

According to our baseline, these two smells are correctly identified. However, the step also exhibits \textsc{Unverified Action}: the instruction asks the tester to execute a command and inspect its output, yet the surrounding steps do not provide an explicit, dedicated verification (i.e., a clear \texttt{<dd>} check) that confirms the intended system behavior. In this case, \gemini{} does not flag \textsc{Unverified Action}, likely because the expected value is embedded inline rather than expressed as a standalone verification step, and because the subsequent step is also an action, which further obscures the missing verification structure.

A non-negligible portion of the disagreements with our baseline involved \textsc{Ambiguous Test}. For example, consider again the test case \texttt{1482\_Wireless Networking/hwn-001} and the following action step:

\begin{stepexample}
<dt>After approximately 30 seconds, open the network manager</dt>
\end{stepexample}

\gemini{} does not flag any smell for this step. In contrast, our domain specialist labeled it as \textsc{Ambiguous Test}, because the waiting time (``approximately 30 seconds'') is not precisely defined and may introduce variability in outcomes across executions or environments. This example also illustrates that Ambiguous Test can be inherently subjective: depending on the context, ``approximately'' may be acceptable operational guidance, yet it can also be interpreted as underspecification when timing differences can affect the observed behavior. Consequently, in borderline cases like this, it is not always straightforward to determine whether a step should be considered ambiguous.

\subsection{\gpt{}}
\label{sec:gpt}

Several other LLMs are available for this task. Among them, \gpt{} has shown strong performance in community-driven evaluations such as LLM Arena~\cite{llm-arena}. In February 2026, we manually executed the same prompt described in Section~\ref{s:methodology-prompt} using \gpt{} in thinking mode, with one temporary conversation per query to avoid cross-query contamination. We evaluated \numStepsManual{} test steps sampled from \numTestsManual{} test cases, for which the first author had previously established a manual baseline. Using this baseline, our \gpt{}-based approach achieves \precisionGPT{} precision and \recallGPT{} recall, with an F$_1$ score of \fscoreGPT{}. Overall, its performance is inferior to \gemini{}.

Overall, \gpt{} correctly detected most smell categories, but it missed some cases. 

We also revisited the two representative disagreement cases discussed in Section~\ref{sec:manual-analysis}, where \gemini{} failed to reproduce the manual baseline. Interestingly, \gpt{} correctly identified the smells missed by \gemini{} in both cases.

In test case \texttt{image/1303\_Live Session (Test-case Live Session Start)}\footnote{\href{https://github.com/easy-software-ufal/manual-test-sensei/blob/main/testcases/image/1303_Live\%20Session}{Manual test case \#1303 (image)}}, \gpt{} matched the baseline by flagging the step as ambiguous, whereas \gemini{} reported no smell. The analyzed step is shown next:

\begin{stepexample}
<dt>Boot up the image</dt>
\end{stepexample}

\gpt{} identifies the expression ``the image'' as under-specified:

\begin{geminioutput}
\textbf{Smells:} Ambiguous Test\\
\textbf{Explanation:} Target is under-specified: 'the image' is not uniquely identified (no name/ID/version/path).
\end{geminioutput}

By contrast, \gemini{} does not flag the step:

\begin{geminioutput}
\textbf{Smells:} None\\
\textbf{Explanation:} No clear smell indicators in the step given its field.
\end{geminioutput}


A similar situation occurred in \texttt{hardware/1482\_Wireless Networking}\footnote{\href{https://github.com/easy-software-ufal/manual-test-sensei/blob/main/testcases/hardware/1482_Wireless\%20Networking}{Manual test case \#1482 (hardware)}}, where \gpt{} correctly identified ambiguity in a step that \gemini{} considered smell-free. The analyzed step is shown next:

\begin{stepexample}
<dt>After approximately 30 seconds, open the network manager</dt>
\end{stepexample}

\gpt{} identifies the phrase ``approximately 30 seconds'' as vague timing, which may hinder consistent test execution:

\begin{geminioutput}
\textbf{Smells:} Ambiguous Test\\
\textbf{Explanation:} Uses vague timing ``approximately 30 seconds'', which is under-specified for consistent execution.
\end{geminioutput}

By contrast, \gemini{} does not flag the step:

\begin{geminioutput}
\textbf{Smells:} None\\
\textbf{Explanation:} No clear smell indicators in the step given its field.
\end{geminioutput}

Together, these examples suggest that, although \gemini{} achieves better aggregate performance, \gpt{} may be more sensitive to subtle ambiguity and more closely aligned with human judgment in borderline cases.

\subsection{SLMs}
\label{sec:slms}

In our prior work~\cite{keila-sbes-2025}, we evaluated \gemma{}~\cite{team2025gemma}, \llama{}~\cite{touvron2023llama}, and \phimodel{}~\cite{abdin2024phi} on 261 individual test steps extracted from 143 real-world Ubuntu manual test cases, targeting seven natural language test smells. Among the evaluated SLMs, \phimodel{} achieved the best performance, reaching a \(pass@2\) of 97\% for detecting steps containing test smells, while \gemma{} and \llama{} achieved around 91\%. Beyond accuracy, the results indicated that SLMs can provide a low-cost, concept-driven alternative to rule-heavy approaches, enabling smell identification without extensive syntactic analysis and offering practical advantages such as local deployment and improved data privacy.

In the present work, we extend this investigation in three key ways. First, we move from step-level inputs to analyzing complete test cases, which better reflects how manual tests are authored and reviewed, but also increases input length and contextual dependencies, potentially challenging smaller models. Second, we strengthen the prompting strategy by embedding the full set of smell definitions and explicit identification rules in the prompt, and by enforcing strict evidence-grounded explanations. Compared to our earlier SLM setting, this reduces the likelihood of guessing and encourages the model to align predictions with concrete textual cues and with the action versus verification roles implied by the Ubuntu \texttt{<dl>}/\texttt{<dt>}/\texttt{<dd>} structure. Third, we scale the analysis to \numTestCases{} complete Ubuntu test cases, allowing us to better characterize smell prevalence in practice. While this LLM-based setting introduces higher inference cost and requires sending data to an external API, our results show that \gemini{} achieves excellent performance even with \(pass@1\), and produces concise explanations that can support practitioners in improving the clarity and consistency of manual test cases.

Analyzing each test step in isolation, as in our prior work~\cite{keila-sbes-2025}, can lead to false positives because some smells depend on information that is only available at the test-case level. In such situations, considering the surrounding steps provides essential context. For example, consider the test case
\texttt{hardware/1467\_External Microphone}.\footnote{\href{https://github.com/easy-software-ufal/manual-test-sensei/blob/main/testcases/hardware/1467_External\%20Microphone}{Manual test case \#1467 (hardware)}} and the following action step:

\begin{stepexample}
<dt>Connect the microphone to the computer. (This can include a microphone on a headset)</dt>
\end{stepexample}

\noindent The annotation produced by \gemini{} is as follows:

\begin{geminioutput}
\textbf{Smells:} Unverified Action\\
\textbf{Explanation:} Action step `Connect' is not followed by a verification step and contains no verification language.
\end{geminioutput}

Under a strictly step-level analysis, our previous approach would likely flag \textsc{Ambiguous Test}, since the referenced commands are not listed within the step itself. However, when inspecting the full test case, the missing details are provided in subsequent steps: the test case enumerates seven concrete commands to execute. This example highlights a key trade-off. Step-level processing can be useful for models with limited context windows~\cite{context-window}, but it may also overlook cross-step dependencies and thereby produce incorrect smell assignments when critical information is distributed across the test case.

Overall, the comparison highlights a trade-off between SLMs and LLMs. SLMs offer lower inference cost, easier local deployment, and stronger privacy and control guarantees, but they may be more sensitive to prompt length and long-range context when the input grows from isolated steps to complete test cases. In contrast, LLMs such as \gemini{} tend to handle longer inputs and richer contextual dependencies more reliably and can generate more consistent, evidence-grounded explanations, but at the expense of higher runtime cost and reliance on external APIs, which may raise data governance and reproducibility concerns.

\subsection{Manual Test Sensei}
\label{sec:related-sensei}

Soares~\emph{et al.}~\cite{SoaresAORGSMSFB23} proposed Manual Test Sensei, an NLP-based tool to detect natural language test smells in manual test descriptions, including the smell types considered in this article. Their approach implements a set of hand-crafted identification rules using Python and spaCy~\cite{spacy}, relying primarily on lexical and shallow linguistic cues. In practice, each smell is associated with a list of indicator terms and patterns (e.g., specific keywords, verb forms, or markers of optionality and subjectivity), and the tool flags a step when these indicators are present.

In contrast, we adopt an LLM-based approach that supports richer semantic reasoning beyond surface-level keyword matching. Our work investigates a model-based setting in which \gemini{} performs smell detection by considering the full step text and its role in the test case structure, relying on explicit smell definitions and identification rules embedded in the prompt rather than fixed indicator lists. This design enables more context-sensitive judgments and can reduce false positives caused by purely lexical cues.

We used the results of \sensei{} run on Ubuntu test cases~\cite{SoaresAORGSMSFB23}.\footnote{\url{https://github.com/easy-software-ufal/manual-test-sensei/blob/main/ManualTestSensei/results-20230329-125316-en_core_web_lg.csv}} We evaluated \numStepsManual{} test steps sampled from \numTestsManual{} test cases, for which the first author had previously established a manual baseline. Using this baseline, \sensei{} achieves~\precisionSensei{} precision and~\recallSensei{} recall, with an F$_1$ score of~\fscoreSensei{}. Overall, its performance is inferior to \gemini{} and \gpt{}.

In some steps, both approaches agree. For example, in the Ubuntu test case
\texttt{packages/1650\_hardinfo}\footnote{\href{https://github.com/easy-software-ufal/manual-test-sensei/blob/main/testcases/packages/1650_hardinfo}{Manual test case \#1650 (packages)}},
the following action step appears:
\begin{stepexample}
<dt>Enter the name of the file and Click on "Save" button</dt>
\end{stepexample}
Both Manual Test Sensei and our \gemini{}-based approach correctly flag this step as Eager Action, since it bundles two distinct actions (entering the file name and clicking the ``Save'' button) into a single test step.

However, consider the test case
\texttt{image/1337\_Install (Default)}\footnote{\href{https://github.com/easy-software-ufal/manual-test-sensei/blob/main/testcases/image/1337_Install\%20(Default)}{Manual test case \#1337 (image)}}
and the following action step:
\begin{stepexample}
<dt>At "Write changes to disks", verify that everything is right and select YES</dt>
\end{stepexample}

\noindent The annotation produced by \gemini{} is as follows:

\begin{geminioutput}
\textbf{Smells:} Ambiguous Test; Eager Action; Misplaced Verification\\
\textbf{Explanation:} Contains indefinite pronoun `everything' and subjective adjective `right'. Combines multiple operations: `verify' and `select'. Uses verification verb `verify' in the action field.
\end{geminioutput}
Manual Test Sensei correctly detects \textsc{Eager Action} and \textsc{Misplaced Verification}. It also flags \textsc{Unverified Action} because there is no dedicated \texttt{<dd>} verification following this \texttt{<dt>}. In this particular case, however, the verification is embedded within the action step itself; thus, the absence of a separate \texttt{<dd>} is not necessarily incorrect, but rather reflects a redundant (or mixed) formulation. Finally, Manual Test Sensei does not identify \textsc{Ambiguous Test} in this step, whereas \gemini{} flags ambiguity due to the subjective expression ``everything is right,'' which lacks an objective acceptance criterion.

{Overall, \gemini{} attains the highest exact-match count (87/\amostraManual{}): 51 steps are correctly labeled by all three tools, 18 are shared exclusively with \gpt{}, 10 are shared exclusively with Sensei, and 8 are uniquely correct. \gpt{} is correct in 72/\amostraManual{} steps (51 shared by all, 18 shared exclusively with \gemini{}, 1 shared exclusively with Sensei, and 2 uniquely correct). Sensei is correct in 63/\amostraManual{} steps (51 shared by all, 10 shared exclusively with \gemini{}, 1 shared exclusively with \gpt{}, and 1 uniquely correct). Importantly, the tools are complementary: at least one tool matches the baseline in 91/\amostraManual{} steps, leaving 12 steps where none of the three tools reproduces the baseline exactly.

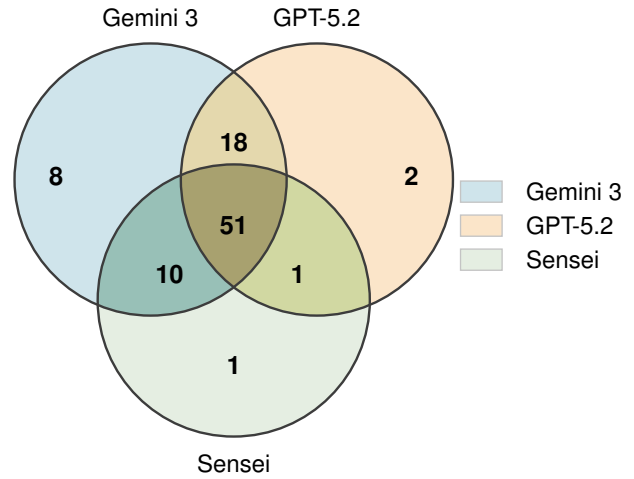
\begin{figure}[t]
  \centering
  \begin{tikzpicture}[font=\small\sffamily]

    \def\r{1.8}
    \coordinate (A) at (-1.1,0.7);   
    \coordinate (B) at ( 1.1,0.7);   
    \coordinate (C) at ( 0.0,-0.9);  

    \definecolor{vennBlue}{RGB}{95,165,190}
    \definecolor{vennOrange}{RGB}{242,171,76}
    \definecolor{vennGreen}{RGB}{170,200,160}
    \definecolor{vennStroke}{RGB}{60,60,60}

    \definecolor{overAB}{RGB}{232,218,170}   
    \definecolor{overAC}{RGB}{140,185,175}   
    \definecolor{overBC}{RGB}{205,215,155}   
    \definecolor{overABC}{RGB}{165,155,105}  

    \tikzset{
      vennOutline/.style={draw=vennStroke, line width=0.9pt},
      vennLabel/.style={font=\small\sffamily},
      vennCount/.style={font=\bfseries\sffamily, text=black}
    }

    \fill[vennBlue,   opacity=0.30] (A) circle (\r);
    \fill[vennOrange, opacity=0.30] (B) circle (\r);
    \fill[vennGreen,  opacity=0.30] (C) circle (\r);

    \begin{scope}
      \clip (A) circle (\r);
      \fill[overAB, opacity=0.78] (B) circle (\r);
    \end{scope}

    \begin{scope}
      \clip (A) circle (\r);
      \fill[overAC, opacity=0.78] (C) circle (\r);
    \end{scope}

    \begin{scope}
      \clip (B) circle (\r);
      \fill[overBC, opacity=0.78] (C) circle (\r);
    \end{scope}

    \begin{scope}
      \clip (A) circle (\r);
      \clip (B) circle (\r);
      \fill[overABC, opacity=0.88] (C) circle (\r);
    \end{scope}

    \draw[vennOutline] (A) circle (\r);
    \draw[vennOutline] (B) circle (\r);
    \draw[vennOutline] (C) circle (\r);

    \node[vennLabel] at ($(A)+(0,\r+0.35)$) {Gemini 3};
    \node[vennLabel] at ($(B)+(0,\r+0.35)$) {GPT-5.2};
    \node[vennLabel] at ($(C)+(0,-\r-0.35)$) {Sensei};

    \node[vennCount] at (-2.35, 0.75) {8};  
    \node[vennCount] at ( 2.35, 0.75) {2};   
    \node[vennCount] at ( 0.00,-1.75) {1};   

    \node[vennCount] at ( 0.00, 1.2) {18};  
    \node[vennCount] at (-0.85,-0.55) {10};  
    \node[vennCount] at ( 0.85,-0.55) {1};   
    \node[vennCount] at ( 0.00, 0.10) {51};  

    \begin{scope}[shift={(3.0,0.40)}]
      \fill[vennBlue,   opacity=0.30, draw=vennStroke, line width=0.5pt] (0,0) rectangle (0.55,0.28);
      \node[anchor=west] at (0.75,0.14) {Gemini 3};

      \fill[vennOrange, opacity=0.30, draw=vennStroke, line width=0.5pt] (0,-0.45) rectangle (0.55,-0.17);
      \node[anchor=west] at (0.75,-0.31) {GPT-5.2};

      \fill[vennGreen,  opacity=0.30, draw=vennStroke, line width=0.5pt] (0,-0.90) rectangle (0.55,-0.62);
      \node[anchor=west] at (0.75,-0.76) {Sensei};
    \end{scope}

  \end{tikzpicture}
  \caption{Venn diagram ($N=\amostraManual{}$) of steps where each tool exactly matches the baseline smell set.}
  \label{fig:venn-acerto-exato}
\end{figure}

\subsection{Threats to Validity}
\label{sec:threats}

This study is subject to threats to validity that may affect our results and interpretations~\cite{threats-llms-icse-nier-2024}. We discuss the main threats and the mitigation actions adopted.

\subsubsection{Internal Validity}
Our evaluation relies on manual annotations for a subset of the analyzed test cases. This process is inherently error-prone and may introduce subjectivity, since some smell definitions can leave room for interpretation and different annotators may disagree on borderline cases. To mitigate this threat, we base labeling on explicit definitions and identification rules, which are also embedded in the prompt, and we require the model to justify predictions by citing concrete textual triggers present in the step. In addition, the manually inspected sample was reviewed to reduce labeling mistakes and to avoid misinterpretations when assessing model outputs and explanations.

\subsubsection{Construct Validity}
We operationalize smell detection as a multi-label classification problem at the step-level within a whole-test-case context. Although we analyze complete test cases, our ground truth is established by labeling individual steps according to the catalogued smells, and our metrics (precision, recall, and F$_1$ score) measure agreement with these labels rather than downstream impacts such as execution time, fault detection effectiveness, or maintenance effort. Moreover, explanation usefulness is assessed qualitatively by checking whether explanations cite evidence aligned with the smell definition, which remains a proxy for practical usefulness rather than a direct measure of developer benefit.

Another threat concerns the prompting strategy adopted in this study. The reported results depend on the specific combination of smell definitions, identification rules, examples, and meta-prompting instructions used to construct the final prompt. Alternative prompt formulations could lead to different performance levels. To facilitate replication and future comparisons, we provide the complete prompt used in our study. Investigating prompt sensitivity remains an important direction for future work.

\subsubsection{External Validity}
Our dataset is drawn from Ubuntu manual test cases, which follow a specific writing style and a structured \texttt{<dl>}/\texttt{<dt>}/\texttt{<dd>} representation. While Ubuntu is a widely used open-source system with a mature QA process, the prevalence of smells and the model performance may differ in other projects, domains, or manual testing formats. In addition, our evaluation is limited to a single contemporary model available at the time of the study (\gemini{}). Therefore, the results may not generalize to other LLMs or to smaller open models without further investigation.

Our results may also not fully generalize over time because LLM behavior can change due to model updates and API-level defaults, and outputs may vary under non-deterministic decoding. To support future replications, we report the execution period (January and February 2026) and adopt a fixed prompting strategy with strict output constraints. Even so, performance and explanations may differ in later versions of the model.

\subsubsection{Conclusion Validity}
Our conclusions about the effectiveness of \gemini{} are based on the test cases analyzed and on a manually inspected sample used to compute performance metrics. Although we observe frequent smell occurrences and encouraging detection performance, broader conclusions require additional evaluations on larger samples, more projects, and alternative test repositories. We therefore interpret the findings as evidence of practical promise rather than as definitive proof of general applicability.

\section{Related Work}
\label{sec:related}

Hauptmann~\emph{et al.}~\cite{Hauptmann2013} argue that natural language test cases, which are widely used in system testing, are often written without following established best practices and therefore become difficult to maintain, hard to understand, and inefficient to execute. Inspired by the established notions of code smells and test smells, the authors introduce the concept of Natural Language Test Smells to characterize recurring quality problems in manual test cases written in natural language. To investigate the prevalence of these issues in practice, they conduct an empirical study on more than 2,800 test cases from seven industrial test suites, examining the occurrence of Natural Language Test Smells in real-world system-testing artifacts.

Soares~\emph{et al.}~\cite{SoaresAORGSMSFB23} present an empirical study on test smells in manual tests written in natural language, motivated by the observation that poorly described manual tests can suffer from issues analogous to those found in automated tests, potentially harming maintainability, coverage, and reliability. To address the limited knowledge about the types, frequency, and impact of such smells, the authors aim to contribute a dedicated catalog for manual test artifacts. They follow a two-fold strategy. First, they conduct an exploratory study on manual tests from three systems (Ubuntu, the Brazilian Electronic Voting Machine, and the user interface of a major smartphone manufacturer), and propose a catalog of eight natural language test smells along with identification rules based on syntactic and morphological text analysis, which they validate with 24 in-company test engineers. Second, they implement an NLP-based tool derived from these rules to automatically analyze the tests of the studied systems and validate its results. Their findings show frequent occurrences of the eight smells in practice; practitioners largely agree with the proposed definitions and examples (80.7\% agreement), and the tool achieves high detection effectiveness (precision 92\%, recall 95\%, F$_1$ score 93.5\%), reporting 13,169 smell occurrences across the analyzed systems. The study concludes that a dedicated catalog and NLP-based detection strategies can reduce the effort of analyzing manual tests, including in multilingual contexts.

Aranda~\emph{et al.}~\cite{Aranda-ease2024} address the problem of test smells in natural language manual tests, noting that such smells can hinder testing activities by reducing maintainability, inducing non-deterministic behavior, and leading to incomplete verification. While prior work has catalogued smells in natural language tests, the authors highlight a gap in systematic and automated approaches for their removal. To fill this gap, they introduce (i) a catalog of transformations designed to remove seven natural language test smells and (ii) a companion tool that applies these transformations using Natural Language Processing techniques. Their goal is to improve the quality and reliability of natural language tests during software development. The paper evaluates these contributions through a two-fold empirical strategy. First, a survey with 15 software testing professionals assesses the acceptance and perceived usefulness of the proposed transformations. Second, an empirical study evaluates the automated tool by applying it to real-world test cases from the Ubuntu operating system. The results suggest that practitioners consider the transformations valuable, and the tool achieves a promising level of effectiveness, reporting an F-measure of 83.70\%.

Soares~\emph{et al.}~\cite{emse-2025} investigate the impact of test smells in natural language manual test descriptions, a topic that has received substantially less attention than smells in automated tests. Motivated by the observation that such smells may harm maintainability, introduce non-deterministic execution, and lead to incomplete verification, the authors note that prior work has mostly focused on cataloging smells, with limited empirical evidence about their effects on test effectiveness. To address this gap, they conduct a controlled experiment with 30 participants from academia and industry and study two smells, Ambiguous Test and Eager Action. They evaluate whether these smells increase (i) test execution time, (ii) the number of steps or screens required to complete the tests (screen flow), and (iii) divergence in participants' perceptions of test success. Their results show that Ambiguous Test can increase execution time by up to five times and screen flow by up to seven times, whereas Eager Action does not increase execution time or screen flow when the grouped actions are dependent on one another. The study concludes by highlighting the need for better-designed manual test descriptions to improve clarity, consistency, and execution efficiency.

Beyond test cases, related efforts also use NLP to identify quality issues in other natural language artifacts. Viezaga~\emph{et al.}~\cite{veizaga2024}, for example, proposed an automated approach to detect ambiguous or problematic linguistic patterns in requirements, and argued that contextualized rephrasing suggestions can assist analysts in improving specification quality. 

In a similar spirit, Rajkovic and Enoiu~\cite{rajkovic2022} developed NALABS to detect quality issues in requirements and test specifications written in natural language. Their approach relies on keyword-based indicators to compute metrics such as vagueness, referencability, optionality, subjectivity, and fragility, illustrating a representative class of rule-oriented methods for smell-like phenomena.

Other works have also explored test smells and related quality issues through the lens of language models. Yang~\emph{et al.}~\cite{yang2024} discussed challenges and systematic methods for exploring new smell types and combined multiple detection techniques to identify diverse smells. From a broader perspective, 

Lucas~\emph{et al.}~\cite{lucas2024} investigated the use of LLMs to detect a large set of test smells, emphasizing that these models can identify issues and often provide improvement suggestions. More generally, LLM-based quality analysis has been applied to other software artifacts as well. 

Wu~\emph{et al.}~\cite{wu2024} proposed \textit{iSMELL}, which integrates LLMs with specialized tools to detect and refactor code smells, dynamically selecting tools to improve scalability and performance. Although these studies focus on different artifacts and smell taxonomies, they collectively motivate the use of language models as flexible analyzers that can operate without extensive hand-crafted rule sets.

Bavota~\emph{et al.}~\cite{Bavota2015} conducted a controlled experiment with 20 master’s students to investigate whether test smells hinder source code comprehension during software maintenance. The experiment involved program comprehension tasks performed on test suites with and without test smells, with participant performance measured in terms of correctness and time. The results indicate that several test smells have a strong negative effect on the understandability of both test suites and production code. In particular, test comprehension was found to improve by approximately 30\% in the absence of test smells, reinforcing that these smells can substantially increase the cost of understanding and maintaining test artifacts.

In this article, we build on these foundations by evaluating \gemini{} for detecting natural language test smells in Ubuntu-style manual test cases. In contrast to prior rule-based or keyword-driven approaches, and extending our earlier investigation of SLMs~\cite{keila-sbes-2025}, we assess a contemporary LLM in a whole-test-case setting. Our approach leverages the \texttt{<dl>}/\texttt{<dt>}/\texttt{<dd>} structure to interpret action and verification roles, which is essential for role-dependent smells and for identifying unverified actions. We detected test smells in the \numTestCases{} Ubuntu test cases analyzed, indicating that these issues occur frequently in practice. Moreover, in the manually inspected sample, \gemini{} achieved better performance than the SLMs evaluated in our prior work~\cite{keila-sbes-2025}, and produced concise, evidence-grounded explanations, suggesting that model-based detection can provide actionable support for practitioners aiming to improve the clarity and consistency of manual test cases.

\section{Conclusions}
\label{sec:conclusions}

In this article, we investigated the use of LLMs for detecting natural language test smells in Ubuntu-style manual test cases. We focused on a whole-test-case analysis setting, in which the model analyzes complete manual tests rather than isolated test-case steps. This setting allows the model to exploit contextual information across actions and verifications and better reflects how practitioners inspect manual testing artifacts. We operationalized seven smell types through a structured prompting strategy that leverages the \texttt{<dl>} organization of Ubuntu tests, and we showed that model-generated explanations can support the identification of ambiguous, complex, misplaced, or unverified steps. Overall, our results provide evidence that LLM-based detection can scale test smell analysis beyond rule-based techniques and support quality assessment of manual testing artifacts.

Our findings also highlight clear benefits when moving from SLMs to a contemporary LLM. In particular, \gemini{} consistently outperformed the SLMs evaluated in our earlier study~\cite{keila-sbes-2025}, while producing more actionable explanations that can help practitioners revise test steps for greater clarity and consistency. This performance gap suggests that, at least for this task and dataset, stronger models can better capture subtle linguistic cues and contextual dependencies that are common in manual test descriptions.

At the same time, deploying model-based smell detection in real workflows raises challenges. Inference cost and latency may be non-trivial for large test suites, and model outputs can vary across prompts, decoding settings, and model updates. Moreover, incorrect classifications may introduce noise in review processes if not accompanied by conservative thresholds and human oversight. Despite these constraints, our results are promising and suggest that language models can provide practical assistance for improving manual tests, especially when the goal is to support reviewers and testers rather than to fully replace human judgment.

First, we plan to expand the evaluation to substantially more test cases and to additional manual-testing repositories beyond Ubuntu, including test suites written in other languages, to assess generalization across writing styles, domains, and linguistic contexts. Given the strong performance of current models on our English dataset, we expect this direction to be feasible and potentially effective, particularly for widely supported languages.
Second, we will consider more projects and larger, more diverse test suites to better characterize how smell prevalence and detection performance vary across contexts.
Third, we will incorporate agent-based workflows, following recent approaches in the literature~\cite{rian-agentes-test-smells-2026}, to iteratively inspect test cases, cross-check step-level predictions with test-case-level context, and reconcile borderline cases (e.g., ambiguity versus acceptable flexibility).
Fourth, beyond accuracy-oriented measures, we will evaluate reliability and stability to quantify the consistency of model outputs across repeated runs.
Specifically, we will adopt a set of accuracy and stability metrics inspired by prior work on LLM evaluation~\cite{chen2021evaluating,atil2025nondeterminismdeterministicllmsettings}, which allow us to assess whether a model produces correct results and whether it does so consistently across multiple attempts---an essential property for practical developer-facing tools.
Fifth, and most importantly, we will investigate to what extent models can \emph{refactor} natural language manual tests to remove these smells, building on our prior experience with refactoring automated test cases to eliminate test smells~\cite{rian-agentes-test-smells-2026,SoaresRGAS23,soares2020Refactoring}.
Concretely, we will evaluate model-assisted editing suggestions that (i) split coarse steps to mitigate \textit{Eager Action}, (ii) rewrite vague statements into objectively checkable criteria to mitigate \textit{Ambiguous Test}, (iii) relocate checks between action and verification fields to mitigate \textit{Misplaced Action} and \textit{Misplaced Verification}, (iv) make conditions explicit and structured to mitigate \textit{Conditional Test}, and (v) add missing expected outcomes to mitigate \textit{Unverified Action}.
We will measure not only smell reduction (before/after) but also semantic preservation of the intended procedure, readability, and the amount of human effort required to accept or revise the proposed edits.
Finally, we aim to evaluate additional model families and sizes and to explore fine-tuning strategies that distill the performance of proprietary models into smaller and lighter open models, enabling lower-cost and privacy-preserving deployments without sacrificing accuracy.
Together, these directions can further clarify the practical boundaries and opportunities for scalable, model-assisted quality assurance and refactoring of natural language manual tests.


\subsection*{Acknowledgements}
We thank the anonymous reviewers for their constructive feedback and insightful suggestions, which significantly improved the quality and clarity of this article.
This work was partially supported by CNPq (306026/2026-0, 408040/2025-4, 403719/2024-0), CAPES (88887.313474/2026-00), FAPESQ-PB (268/2025), and FAPEAL grants. This work was partially supported by INES.IA (National Institute of Science and Technology for Software Engineering Based on and for Artificial Intelligence) \url{http://www.ines.org.br}, CNPq grant 408817/2024-0.


\section*{Availability of data and materials}
All artifacts are publicly available online~\cite{artefatos}.

\bibliographystyle{plain}
\balance

\end{document}